\documentclass{ws-ijmpe}
\usepackage{graphicx} 
\begin{document}

\markboth{S. Salur for the CMS Collaboration}{Jet results and jet reconstruction techniques in p+p and their prospects in Pb+Pb collisions in CMS}
\catchline{}{}{}{}{}

\title{JET RESULTS AND JET RECONSTRUCTION TECHNIQUES IN P+P AND THEIR 
PROSPECTS IN PB+PB COLLISIONS IN CMS}

\author{\footnotesize SEVIL SALUR FOR THE CMS COLLABORATION}

\address{University of California at Davis, One Shields Ave.
Physics Department
Davis, CA 95616, USA\\ 
salur@physics.ucdavis.edu}

\maketitle

\begin{history}
\end{history}

\begin{abstract}
Copious production of very energetic jets is expected at the LHC due to the
large increase in collision energy. Jet reconstruction at these high center of mass energies will provide crucial leverage to map out the
QCD evolution of parton energy loss and a unique insight into the nature of the hot QCD matter. This article presents jet reconstruction techniques
and the preliminary jet results in p+p collisions at $\sqrt{s}=7$ TeV collected by the CMS experiment at the LHC.  Jet reconstruction prospects for the heavy ion collisions are also discussed.
\end{abstract}

\section{Introduction}

Jet modification has been observed in high energy nuclear collisions via inclusive particle spectra and di-hadron correlations at RHIC \cite{star1,star2,phenix}. However
these measurements are intrinsically biased towards the jet fragments that interacted least in the medium. A more sensitive measurement of jet quenching can be performed with full jet reconstruction in heavy-ion collisions.  In this article, the preliminary jet results in p+p collisions at $\sqrt{s}=7$ TeV are discussed with jet reconstruction prospects in Pb+Pb collisions at LHC utilizing the CMS experiment. The first of the following sections discusses the jet reconstruction methods in CMS. This section is followed by the techniques 
used to calibrate the jet energy scale and a brief discussion of jet triggering capabilities. Finally the results in p+p and the prospects in heavy ion collisions are reviewed.

\section{Jet Reconstruction in CMS}

The two most important detector requirements for a successful reconstruction of jets in heavy ion collisions is a good energy measurement of particles and an efficient jet trigger. The CMS experiment at the LHC is a multi-purpose detector designed to explore the physics at the TeV energy scale. With high quality electromagnetic and hadronic calorimeters covering a wide range in pseudorapidity for the full azimuthal angle, by design CMS is very well suited to measure hard scattering processes. Sub-detectors such as the high precision silicon tracker with a very good momentum resolution are complimentary to the calorimeters for jet studies. For example the silicon tracker can be used to correct the jet energy for distortions due to the large magnetic field affecting the energy resolution of the low $E_{T}$ jets. All sub-systems are built with high resolution detectors to cope with the high luminosity p+p collisions. Also by design CMS detectors are well suited to study events with large multiplicites resulting from heavy ion collisions.

There are four types of jet reconstruction methods in use by  CMS which differ in the way the sub-detector input particles are characterized.  The first one of these methods is the calorimeter jets which are reconstructed using the calorimeter towers that are combined from energy deposits in hadronic and electromagnetic calorimeter cells. The second method is called jet plus tracks (JPT) and has been developed to correct the calorimeter jets  energy and the direction by utilizing the silicon tracker.   The JPT method improves the $p_{T}$ response and resolution especially in the low $p_{T}$ region where the calorimetric response is poor. The particle flow jets (PFJ) reconstruct jets starting from all stable particles in the event, i.e.,  electrons, muons, photons and charged and neutral hadrons by combining information from all sub-detectors \cite{7}. The last method reconstructs  track jets from  only tracks of charged particles measured in the high precision silicon tracker. Since it is completely independent from the calorimetric measurements, it is used as a cross-check method for the jet results \cite{hinzmann}.

In addition to these four methods, CMS uses several jet algorithms to combine the input particles into jets. The default algorithm for p+p collisions is anti-$k_{T}$ with a resolution parameter of 0.5 \cite{akt}. Other sequential recombination and cone algorithms that are encoded in the $FastJet$ framework and integrated in the CMS software will be explored with the first Pb+Pb collisions starting this November \cite{fastjet,newpap}.

\section{Jet Energy Calibration}

CMS uses a factorized multi-step approach to correct the jet energy \cite{jme07}. The relative ($\eta$) and the absolute ($p_{T}$) correction steps are required to correct for the variation in jet energy response in pseudorapidity and the transverse momentum consecutively.  There are also additional optional corrections that take into account jet response variations  in electromagnetic energy fraction and in flavor (e.g., light quark, b, c or gluon). Corrections due to soft interactions involving spectator partons and offset due to electronic noise and pile up are also optional. The factorized approach allows to study  and optimize each correction separately. For the results shown in this article only required jet corrections are applied to the uncorrected  jet energy as illustrated in the Equation~\ref{eq:corr}. 
\begin{equation}\label{eq:corr}
Corrected\;Jet\; Energy =(Raw Jet \;Energy - offset)\times C(rel:\eta)\times C(abs:p_{T})
\end{equation}
In addition to utilizing Monte Carlo simulations such as Pythia and full detector simulations based on GEANT, these corrections can be estimated by data driven methods with in-situ collider data as performed by the Tevatron experiments \cite{insitu,insitu2}.   For example with conservation of transverse momentum in a 2 to 2 process, known as $p_{T}$ balance, one of the two final state objects can be measured relative to the other final state object which serves as the reference. CMS employs $p_{T}$ balance in dijet events, $\gamma$ + jet events,  and Z + jet events, to determine jet energy corrections using collider data.  Due to the limited statistics, jet corrections in CMS are for the moment still derived using Monte Carlo truth information from the detector simulations and in-situ corrections are used for verifying the consistency of these simulation results. A systematic uncertainty of $\sim10\%$ on the jet energy is achieved and with the expected increase in statistics, more precise corrections will become available.

\section{Jet trigger capabilities}

\begin{figure}[h]
\begin{minipage}{14pc}
\includegraphics[width=15pc]{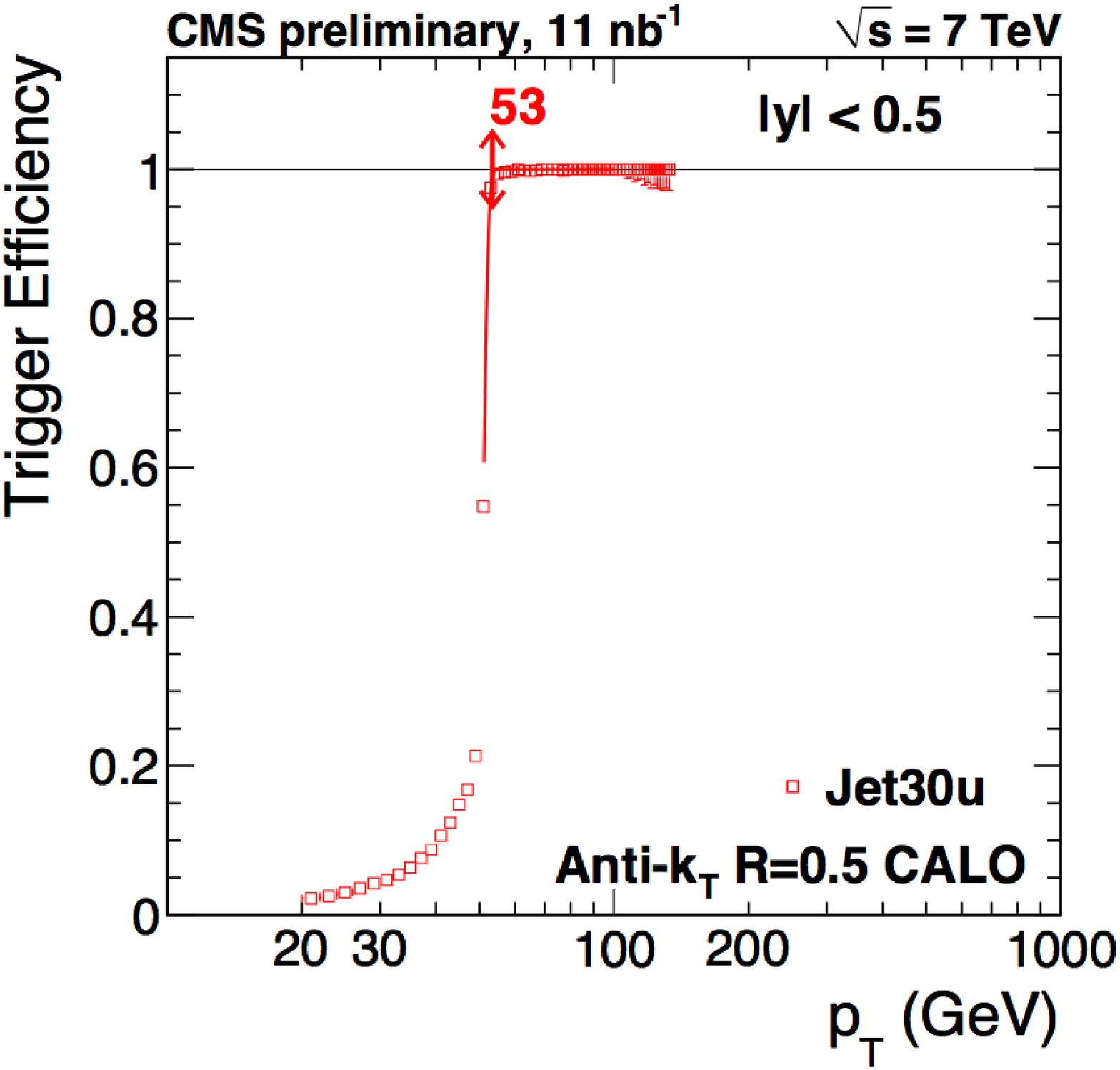}
\caption{\label{fig:eff}   Turn-on 
curve in the central rapidity region versus jet $p_{T}$ for the applied trigger stream for calorimeter
jets. }
\end{minipage}\hfill
\begin{minipage}{14pc}
\includegraphics[width=15pc]{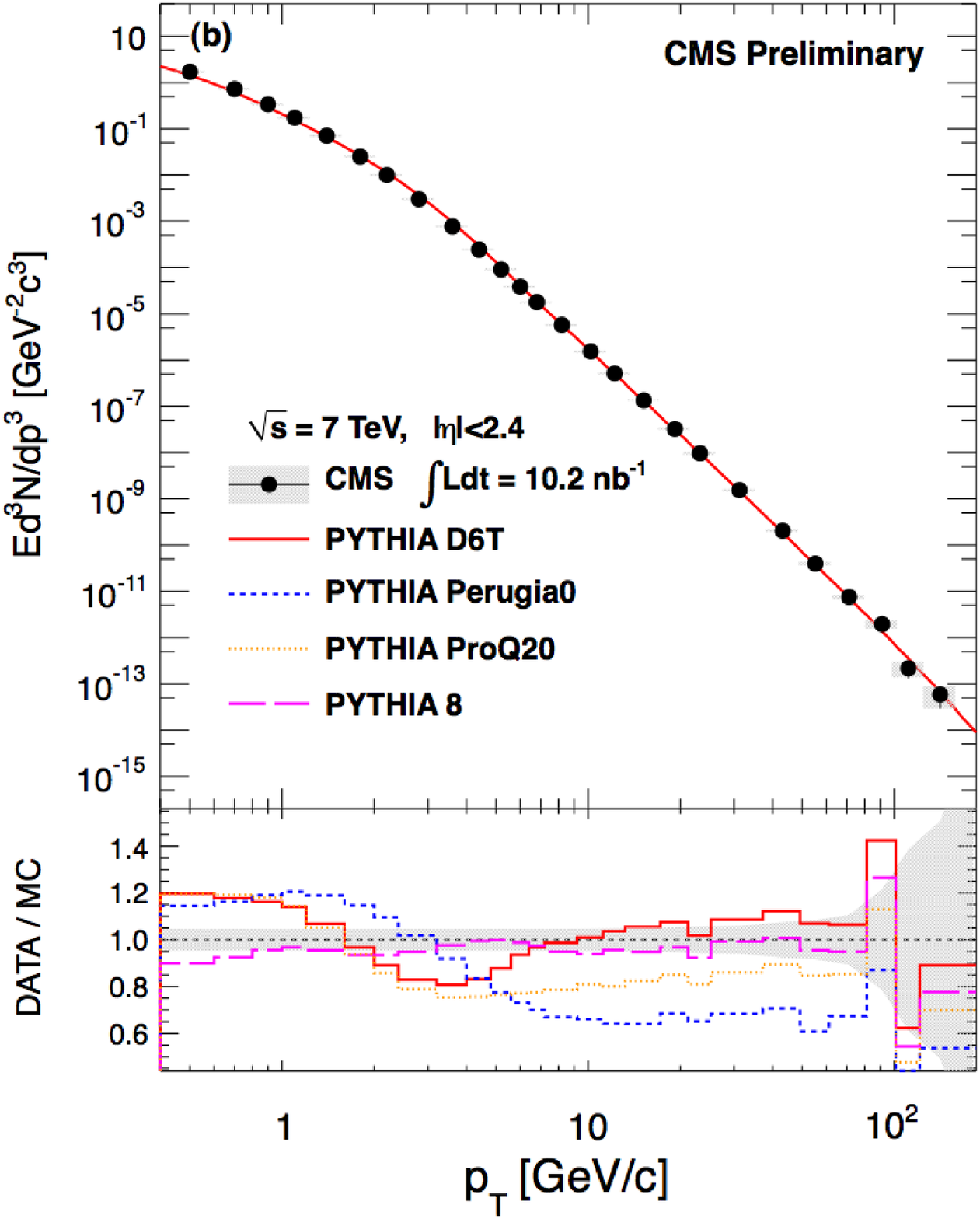}
\caption{\label{fig:chargedspectra}   The invariant differential yield of charged particles at $\sqrt{s}$=7 TeV collisions.  }

\end{minipage} 
\end{figure}

The high readout rate of the CMS data acquisition system allows inspection of all events by a high level trigger (HLT). The jet reconstruction algorithms are fast enough and fit into the HLT time budget to be used to decide for an event trigger decision.  As an example, the Figure~\ref{fig:eff} shows the turn-on curve of the single-jet triggers for calorimeter jets which is estimated by taking the ratio of the single-jet trigger to the minimum bias trigger \cite{1,4,6}.  
Calorimeter jets use a single-jet trigger with 30 GeV uncorrected energy and require that each considered jet have a corrected $p_{T}$ above 50 GeV.  The markers give the estimated turn-on point, after which the trigger is more than 99\% efficient. 

To extend the statistical reach of the measurements and enhance yields at high $p_{T}$,  calorimeter based HLT  jet triggers are employed to measure the charged particle transverse spectra \cite{12}. The invariant differential yield of charged particles from $\sqrt{s}=7$ TeV p+p collisions  are presented in Figure~\ref{fig:chargedspectra}. This spectrum is compared to different PYTHIA tunes over the full $p_{T}$ range. The ratio of the measured CMS measurements to the PYTHIA tunes are presented in the lower inset. The extended kinematic reach is due to the large hard scattering cross-sections at of p+p collisions at $\sqrt{s}=7$ TeV, the large full azimuthal acceptance of the silicon tracker and most importantly the excellent HLT triggering capabilities.

\section{Results in p+p}

 \begin{figure}[h]
\begin{minipage}{15pc}
\includegraphics[width=15pc]{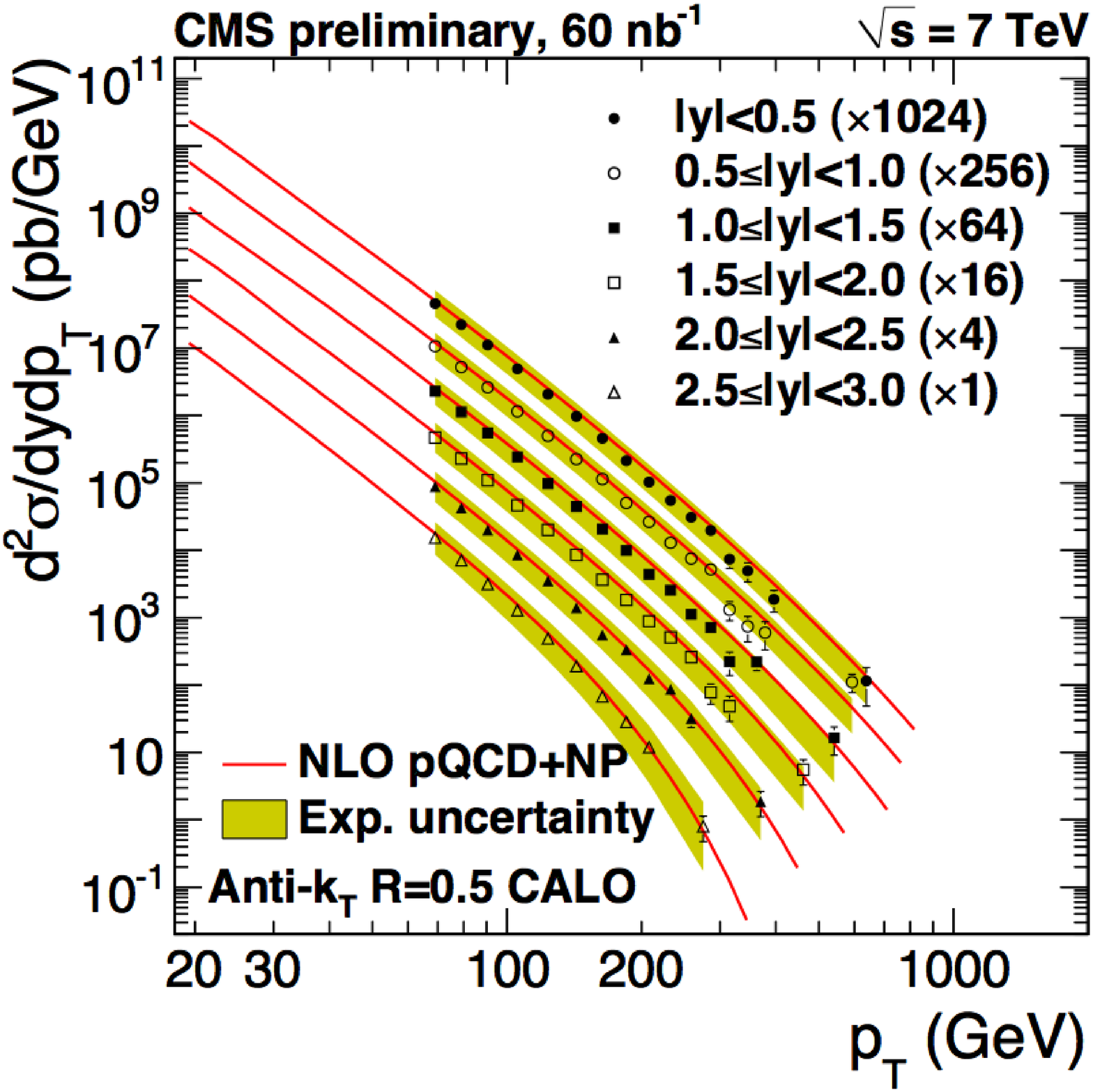}
\end{minipage}\hfill
\begin{minipage}{15pc}
\includegraphics[width=15pc]{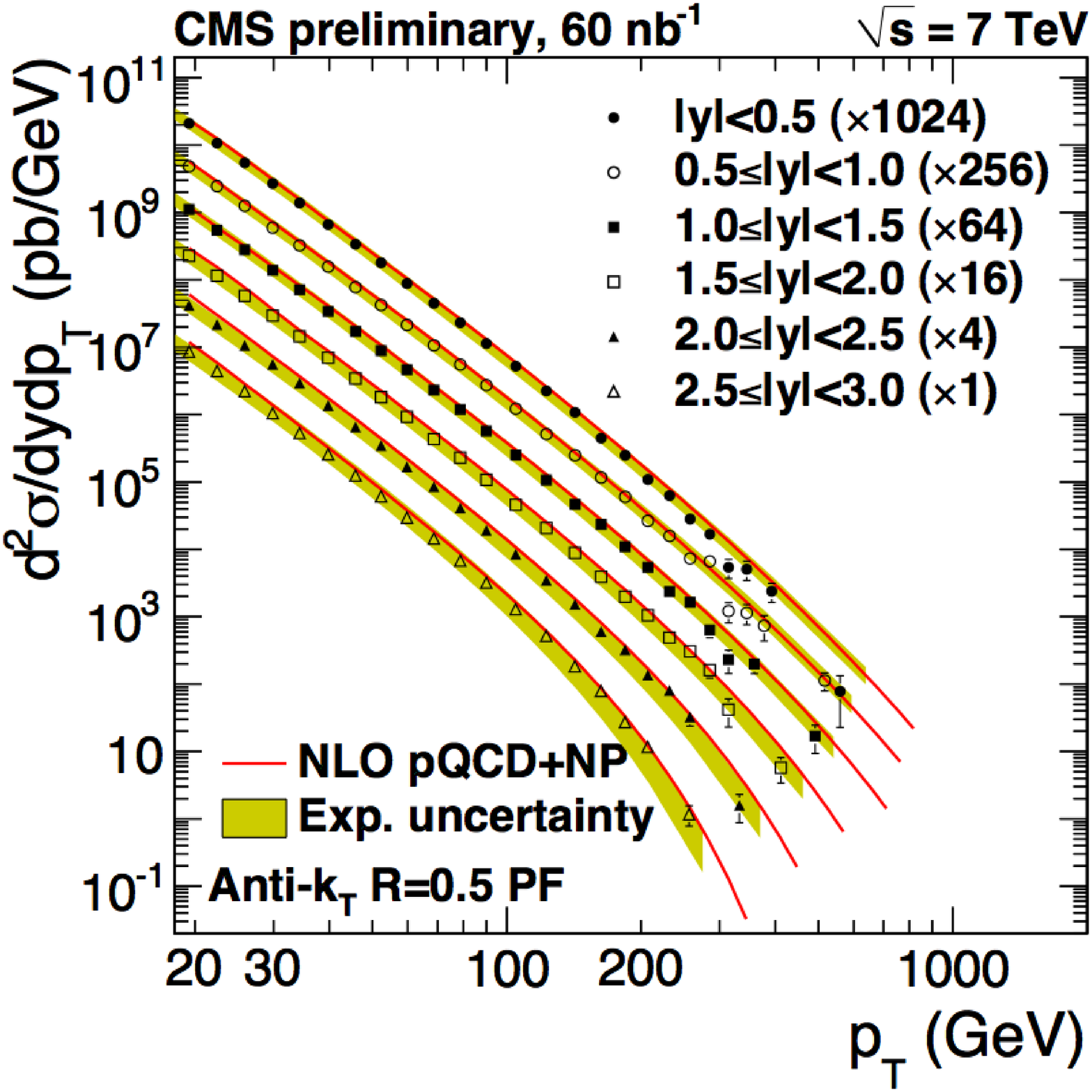}
\end{minipage}

\caption{\label{fig:jetspectra}  Comparison between the corrected measured jet $p_{T}$ spectra and the theory predictions for calorimeter jets on the left and PFJ on the right. }
\end{figure}

The corrected jet $p_{T}$ spectra are presented in Figure~\ref{fig:jetspectra} for calorimeter jets and PFJ from 7 TeV p+p collisions for an integrated luminosity of about 60 n$b^{-1}$. 
The systematic uncertainties are shown as yellow bands and NLO predictions are shown as solid red lines \cite{1}.  For better visibility the spectra for the given rapidity regions are multiplied by arbitrary factors as indicated in the legend. PFJ  extends the kinematic range to much lower jet $p_{T}$ due to the improved jet resolution. The jet results agree with the theoretical prediction and with each other within the uncertainties of the measurements. The high level triggering capability on jets in CMS extends the kinematic reach of measured inclusive jets to large transverse momenta and efficiently composes the whole spectrum down to very low $p_{T}$ by utilizing multiple trigger streams.  Many jet observables are explored with CMS due to this large kinematic reach. Some examples are search for di-jet resonances in the di-jet mass distributions \cite{9}, hadronic event shapes \cite{3}, dijet azimuthal decorrelations and angular distributions \cite{5}, and the underlying event studies \cite{10}.

\section{Prospects for Heavy Ion Collisions}

Jet energy contamination from the underlying heavy ion background will be corrected with a three step procedure.  The first step is measuring the jet area with collinear and infrared safe jet algorithms that are insensitive to splitting and soft radiation \cite{safe}. An active area of each jet is estimated by filling an event with many very soft particles and then counting how many are clustered into a given jet. The second step is measuring the diffuse noise ($< p_T >$ per unit area in the remainder of the event) and noise fluctuations. These fluctuations in the background can distort the jet spectrum towards larger $p_{T}$ which can be corrected through an unfolding procedure (i.e., deconvolution). So the final step is the deconvolution of the signal from the background using parameters that are extracted from measurable quantities.

Relative and absolute correction factors can be initially extracted from Monte Carlo as in the p+p case.  But since parton energy loss modifies the relation between final state particles and the initial parton,  Monte Carlos without a precise description of the jet quenching effects cannot be adequate to estimate the required correction factors. Instead, measurements that are data driven and easily identifiable in a high multiplicity environment such as $\gamma$ + jet, Z + jet, di-jet, jet shapes and fragmentation functions in di-jets can be used to minimize the uncertainty of the jet energy scale calibrations \cite{gunther}. 

One of the channels for the jet tomography of the QGP that is inaccessible at RHIC due to the experimental set-ups or the kinematic limits of RHIC is the Z + jet channel. This channel has been used in Tevatron to calibrate the jet energy in p+p collisions. In heavy ion collisions, Z gauge bosons can again be used as a calibration probe of in-medium parton energy loss as they carry no color charge and leave the medium unattenuated. The short production time of Z bosons and subsequent decay guarantees that production and decay happen before or during the formation of the deconfined QCD medium. This measurement is feasible with a yield of $\sim$ 2000 Z+jet events that decay into dimuons within the CMS acceptance from the expected one month of $\sqrt{s_{NN}}=5.5$ TeV Pb+Pb run. 

\section{Conclusion}

The QCD program with jet reconstruction has started with an amazing performance with the first 7 TeV p+p data and the results are already started pushing Tevatron boundaries.  Detector simulations show a good agreement with the measured p+p data affirming a successful detector commissioning. Monte Carlo models tuned to LEP and Tevatron data already work reasonably at LHC for the jet observables.  Multiple jet reconstruction methods combining calorimetry and tracking offer many new handles on reducing experimental systematic uncertainties in p+p and Pb+Pb collisions. To avoid experimental biases a careful choice of jet observables will be used to study medium modification of jet properties in central Pb+Pb collisions. CMS is ready and waiting for the heavy ion collisions. 



\begin{thebibliography}{0}

\bibitem{star1}STAR Collaboration, J. Adams et al., ``Evidence from d + Au measurements for final state suppression of high $p_{T}$ hadrons in Au+Au collisions at RHIC'', Phys. Rev. Lett. 91 (2003) 072304, [arXiv:nucl-ex/0306024].

\bibitem{star2}STAR Collaboration, J. Adams et al., ``Direct observation of dijets in central Au+Au collisions at $\sqrt{s_{NN}}$ = 200-GeV'', Phys. Rev. Lett. 97 (2006) 162301, [arXiv:nucl-ex/0604018].

\bibitem{phenix}PHENIX Collaboration, S. S. Adler et al., ``Common suppression pattern of eta and pi0 mesons at high transverse momentum in Au+Au collisions at $\sqrt {s_{NN}}=200$ GeV'', Phys. Rev. Lett. 96 (2006) 202301, [arXiv: nucl-ex/0601037]. 

\bibitem{7}CMS Collaboration, ``Commissioning of the Particle-Flow reconstruction in Minimum-Bias and Jet Events from pp Collisions at 7 TeV'', CMS PAS PFT-10-002 (2010).

\bibitem{hinzmann} See contribution by A. Hinzmann for the CMS Collaboration in these proceedings.  

 \bibitem{akt}M. Cacciari, G. P. Salam and G. Soyez, ``The anti-kt jet clustering algorithm", JHEP 0804 (2008) 063, [arXiv:0802.1189]. 

 \bibitem{fastjet}M. Cacciari, G. Salam and G. Soyez, http://www.lpthe.jussieu.fr/$\sim$salam/fastjet/. 
 \bibitem{newpap}M. Cacciari, J. Rojo, G. P. Salam, G. Soyez, ``Jet Reconstruction in Heavy Ion Collisions'', [arXiv:1010.1759].

\bibitem{jme07}CMS Collaboration, ``Plans for Jet Energy Corrections at CMS'', CMS PAS  JME-07-002 (2007).

\bibitem{insitu}D0 Collaboration, B. Abbott et al., ``Determination of the absolute jet energy scale in the D0 calorimeters,Ó Nucl. Instrum. Meth. A424 (1999) 352Ð394.
\bibitem{insitu2} CDF Collaboration, A. Bhatti et al.,``Determination of the jet energy scale at the Collider Detector at Fermilab,Ó Nucl. Instrum. Meth. A566 (2006) 375Ð412.

\bibitem{1} CMS Collaboration, ``Measurement of the Inclusive Jet Cross Section in pp Collisions at 7 TeV'', CMS PAS QCD-10-011 (2010).
\bibitem{4} CMS Collaboration, ``Jet Transverse Structure and Momentum Distribution in pp Collisions at 7 TeV'', CMS PAS QCD-10-014 (2010).
\bibitem{6} CMS Collaboration, ``Jet Performance in pp Collisions at $\sqrt{s}$=7 TeV'', CMS PAS JME-10-003 (2010).


\bibitem{12} CMS Collaboration, ``Charged hadron transverse momentum spectra in pp collisions'', CMS PAS QCD-10-008 (2010).

\bibitem{9} CMS Collaboration, ``Search for Dijet Resonances in the Dijet Mass Distribution in pp Collisions at Sqrt(s)=7 TeV'', CMS PAS EXO-10-001 (2010).

\bibitem{3} CMS Collaboration, ``Hadronic Event Shapes in pp Collisions at 7 TeV'', CMS PAS QCD-10-013 (2010).

\bibitem{5} CMS Collaboration, ``Dijet Azimuthal Decorrelations and Angular Distributions in pp Collisions at 7 TeV'', CMS PAS QCD-10-015 (2010).


\bibitem{10} CMS Collaboration, ``The underlying event in proton - proton collisions at 900 GeV'', CMS PAS QCD-10-001 (2010).

\bibitem{safe}M. Cacciari, G. P. Salam and G. Soyez, ``The catchment area of jets", JHEP 0804 (2008) 005  [arXiv:0802.1188].

\bibitem{gunther}Y. Chen, V. Chetluru, Y.J. Lee, C. Loizides, C. Roland, G. Roland, M.B. Tonjes, Y. Yilmaz, and A.S. Yoon, ``Study of photon-tagged jet events in high-energy heavy ion collisions with CMS'', Eur. Phys. J. C 61 (2009) 649. 
 
  
\end{thebibliography}
\end{document}